\documentclass[english,a4paper,11pt]{extarticle}

\usepackage[T1]{fontenc}
\usepackage[utf8]{inputenc}
\usepackage{babel}
\usepackage{array}
\usepackage{mathtext}
\usepackage{textcomp}
\usepackage{amsmath,amssymb}
\usepackage{epstopdf}
\usepackage{graphicx}	
\usepackage{cite}
\usepackage{euscript}   

\usepackage[letterpaper]{geometry}
\newcommand{\comment}[1]{}  

\newcommand{\wl}{{\omega_{\ell}}}
\newcommand{\al}{\alpha}
\newcommand{\LL}{{\EuScript L}}
\newcommand{\MM}{{\EuScript M}}
\newcommand{\define}{\stackrel{def}{=}}
\newcommand{\be}{\begin{equation}}
\newcommand{\ee}{\end{equation}}

\begin{document}

\begin{center}
{\Large \sc Magnetic phase transitions in ultrathin films of different crystal structures}

\phantom{XXX}

{\bf Yury Kirienko\footnote{mailto: yury.kirienko@gmail.com} and 
    Leonid Afremov\footnote{mailto: afremovl@mail.dvgu.ru}}

\phantom{XXX}

{\it Far-Eastern Federal University, Vladivostok, Russia}

\end{center}

\begin{abstract}
The properties of ultrathin films have been studied within the framework of Ising model and the method of random-field interactions.
It is shown that the Curie temperature is inversely proportional to the number of layers. 
Critical exponent $\nu$ has been obtained and it is shown that it does not depend on the type of crystalline lattice.
\end{abstract}
As a characteristic, that allows to evaluate the smallness of magnetic objects, one can take 
the length of the spin-spin correlation $\xi$. It is used to describe the spin fluctuations near the critical temperature.
In the case when the film thickness is less than the correlation length of bulk samples, there appear various size effects.
In particular, the decrease of film thickness can lead to a reduction of phase transition temperature:
according to measurements on films of iron, nickel and cobalt~\cite{Liu1990,Liu1991,Rau1993,Qui1991,
Li1992,Huang1993,Huang1994,Kohlhepp1992}, $T_c$ decreases with decreasing of film thickness $N$ 
(here $N$ is the number of monolayers). 
The experimentally observed dependence of the Curie temperature on the thickness $T_c(N)$ 
can be described by the relation introduced in~\cite{Huang1993,Huang1994}:
\comment{
Так, в работах \cite{Gubin2005,Billas1997magnetism,Billas1996magnetism} 
представлены измерения зависимости удельного магнитного момента 
$\left\langle m\right\rangle$ от количества атомов в наночастицах железа, никеля и кобальта. 
В~этих работах показано, что, по мере уменьшения числа атомов, удельный магнитный момент 
возрастает. При этом, согласно~\cite{Bentivegna1998}, намагниченность насыщения наночастиц $\al$\,--\,$Fe$ 
с размерами 7~нм значительно меньше, чем у массивного образца. Зависимость удельного 
магнитного момента от количества атомов у частиц редкоземельных элементов, в отличие от 
наночастиц $3d$-металлов, имеет обратный ход: момент падает с уменьшением числа 
атомов~\cite{Billas1997magnetism,Jena1992}.}
\begin{equation} \label{GrindEQ__1_} 
    \varepsilon (N) \define
        \frac{T_c\left(N\to\infty\right)-T_c\left(N\right)}{T_c\left(N\to\infty\right)}=c_0{\left(N-\tilde{N}\right)}^{-\lambda },    
\end{equation} 
where $T_c\left(N\to\infty\right) \define T_{\infty}$~-- temperature of bulk sample, 
$c_0$ and $\tilde{N}$~-- constants. The argument $\lambda$ is related to 
the critical exponent of the spin-spin correlation $\nu$ as follows: $\lambda =1/\nu$~\cite{Barber1983}.
And, as was shown in~\cite{Ballentine1989,Ballentine1990,Schulz1994}, the constants and argument 
of the dependency~\eqref{GrindEQ__1_} essentially depend on the type and symmetry of the crystal lattice 
of sprayed film and its substrate.

In this paper we assess the influence of film thickness and its crystal structure on 
the temperature of magnetic phase transition. To solve this problem we use the following model.

\section{Model}

\begin{itemize}
    \item Ultrathin film is composed of $N$ infinite monolayers;
    \item the interaction fields $h$ between spin magnetic moments of the atoms are distributed randomly,
    and the interaction is realized only between the nearest neighbors;
    \item spin magnetic moments are oriented along an axis $Oz$ (approximation of the Ising model) 
    and are equal to $m_0$ in magnitude.
\end{itemize}

According to~\cite{Belokon2001} the distribution function for random interaction fields $h$ on a particle located at the origin 
(in $n$-th monolayer) is defined as:
\be \label{GrindEQ__1a_} 
W_n\left(h\right) = 
    \int{\delta \left(h-\sum_j{h_{nj}\left({\mathbf r}_j,{\mathbf m}_j\right)}\right)
        \prod_j{F_n\left({\mathbf m}_j\right)\delta \left({\mathbf r}_j-{\mathbf r}_{j0}\right)}}
    d{\mathbf r}_jd{\mathbf m}_j,
\ee 
where $\delta (x-x_0)$ -- Dirac delta function, $h_{nj}=h_{nj}\left({\mathbf m}_j,{\mathbf r}_j\right)$~-- 
field produced by atoms with magnetic moments ${\mathbf m}_j$, located at the nodes with coordinates ${\mathbf r}_j$ in $n$-th monolayer.
$F_n\left({\mathbf m}_j\right)$ is the distribution function for the magnetic moments,
which in the approximation of the Ising model for a ferromagnet can be represented as follow:
\be \label{GrindEQ__2_} 
F_n\left({\mathbf m}_j\right)=
    \left({\al}_{nj}\delta \left({\theta}_j\right)+{\beta}_{nj}
    \delta \left({\theta }_j-\pi \right)\right)\delta \left(m_{0j}\right).
\ee 
Here ${\theta}_j$ is the angle between ${\mathbf m}_j$ and $Oz$-axis, ${\al}_{nj}$ and ${\beta}_{nj}$~--
relative probabilities of the spin orientation along (${\theta}_j=0$) and against (${\theta}_j=\pi$) $Oz$-axis respectively;
 $m_{0j}$~-- magnetic moment of the $j$-th magnetic atom. 
 
Now we introduce symmetric notations ${\al}_{+n}\define\al_n$ and ${\al}_{-n}\define 1-{\al}_n$
and consider the set $k_j$ of nearest neighbors of an arbitrary atom numbered as $j$. 
Let denote $z=\dim  k_j$ as its coordination number.
Then let $\Omega_{k_j}$ be the set of all possible products of $\al_{\pm n}$ (where $n \in k_j$),
which contains $z$ various factors with different values of $n$ (therefore $\dim \Omega_{k_j} = 2^z$).
The elements of $\Omega_{k_j}$ are marked as $\wl$:
$\wl \define
    \prod_{n \in k_j}\al_{\pm n} 
    =\prod^{\nu=z}_{(\nu =1;\, l_{\nu }\in k_j) }{{{\al }_{\pm l}}_v}$. In fact, $\ell$ is a number of
binomial permutation of $\al_{\pm n}$ that forms $\Omega$. If $\LL$ is such a set of binomial permutations,
on which $\Omega$ was build, then we can define a similar set $\MM$ of elements $M_{\ell}$:
$M_{\ell} \define
    \sum_{n\in k_j}{\pm m_n}=
    m_0\sum_{n\in k_j}{\pm |{2\al }_n-1|}$.
 
 In the approximation of nearest neighbors and the direct exchange interaction between 
 magnetic atoms, the equation~\eqref{GrindEQ__1a_} can be represented as:
\be \label{GrindEQ__3_} 
    W_n\left(h\right)=
        \sum^{C^{z-n}_n}_{\nu =1}{\sum^{2^n}_{l_{\nu }\in L\left(C^n_z\left(k_j\right)\right)}
        {\omega_{l_{\nu}}\delta \left(h-M_{l_{\nu}}J_{l_{\nu}}\right)}},   
\ee
where $C^n_z\left(k_j\right)$ is a sample of $n$ atoms of the total number of $z$ 
nearest neighbors of $j$-th atom, $J_{l_{\nu }}$ are the constants of exchange interaction 
(that may differ between different monolayers or even inside the same monolayer between different sorts of atoms).

General equation that determines the average relative magnetic moment ${\mu}_n$ in $n$-th monolayer is
\be \label{eq5}
\mu_n = \int\tanh \left( \frac{m_nH}{k_BT}\right)W_n(H)dH.
\ee
Replacing in expressions for $\wl$ and $M_{\ell}$ all ${\al}_{\pm n}$ on their average values 
$\left\langle {\al}_{\pm n}\right\rangle ={(1\pm {\mu}_n)}/{2}$ 
and substituting \eqref{GrindEQ__3_} into \eqref{eq5}, 
one can obtain the equations that determine ${\mu}_n$ in each monolayer:
\be \label{GrindEQ__4_} 
\left\{
\begin{array}{rl}
    \mu_1 &=
        \sum\limits^{z_{1,1}}_{l=0}{C^l_{z_{1,1}}}\frac{{\left(1{+}\mu_1\right)}^l{\left(1{-}\mu_1\right)}^{z_{1,1}-l}}{2^{z_{1,1}}}\times\\
        &\qquad \sum^{z_{1,2}}_{k=0}{C^k_{z_{1,2}}}\frac{{\left(1+{\mu }_2\right)}^k{\left(1-{\mu }_2\right)}^{z_{1,2}-k}}{2^{z_{1,2}}}
        {\tanh  \left(\frac{2\left(l+k\right)-(z_{1,1}+z_{1,2})}{t}\right)\ },\\
    \mu_n &=
        \sum^{z_{n,n}}_{l=0}{C^l_{z_{n,n}}}\frac{{\left(1{+}\mu_n\right)}^l{\left(1{-}\mu_n\right)}^{z_{n,n}-l}}{2^{z_{n,n}}} 
        \sum^{z_{n-1,n}}_{k=0}{C^k_{z_{n-1,n}}}\frac{{\left(1{+}\mu_{n-1}\right)}^k{\left(1{-}\mu_{n-1}\right)}^{z_{n-1,n}-k}}{2^{z_{n-1,n}}}\times\\ 
        &\qquad \sum^{z_{n,n+1}}_{r=0}{C^r_{z_{n,n+1}}}\frac{{\left(1{+}\mu_{n+1}\right)}^r{\left(1{-}\mu_{n+1}\right)}^{z_{n,n+1}-r}}{2^{z_{n,n+1}}} 
        \tanh  \left(\frac{2\left(l+k+r\right)-(z_{n-1,n}+z_{n,n}+z_{n,n+1})}{t}\right)\\
    \mu_N &=
        \sum^{z_{N,N}}_{l=0}{C^l_{z_{N,N}}}\frac{{\left(1{+}\mu_N\right)}^l{\left(1{-}\mu_N\right)}^{z_{N,N}-l}}{2^{z_{N,N}}}\times\\
        &\qquad \sum^{z_{N-1,N}}_{k=0}{C^k_{z_{N-1,N}}}\frac{{\left(1{+}\mu_{N-1}\right)}^k{\left(1{-}\mu_{N-1}\right)}^{z_{N-1,N}-k}}{2^{z_{N-1,N}}} 
        \tanh  \left(\frac{2\left(l+k\right)-(z_{N-1,N}+z_{N,N})}{t}\right),   
\end{array}
\right.
\ee 
where $z_{n,n}$ is the number of nearest neighbors in the $n$-th layer,
$z_{n-1,n}$ is the number of nearest neighbors of the atom in $(n-1)$-th layer, located  in the $n$-th layer;
$i_{nn}={J_{nn}m_n}/{J_{11}}m_1$, $i_{n-1,n}={J_{n-1,n}m_{n-1}}/{J_{11}m_1}$, $i_{n,n+1}={J_{n,n+1}m_{n+1}}/{J_{11}m_1}$, 
$t={k\ T}/{J_{11}}m_1$. Using  \eqref{GrindEQ__4_}, one can study the dependence of 
the average magnetic moment of the film on its temperature and thickness.

\section{The Curie temperature of ultrathin films}
\begin{figure}
    \begin{center}
        \begin{tabular}{rl}
            \includegraphics[width=9cm]{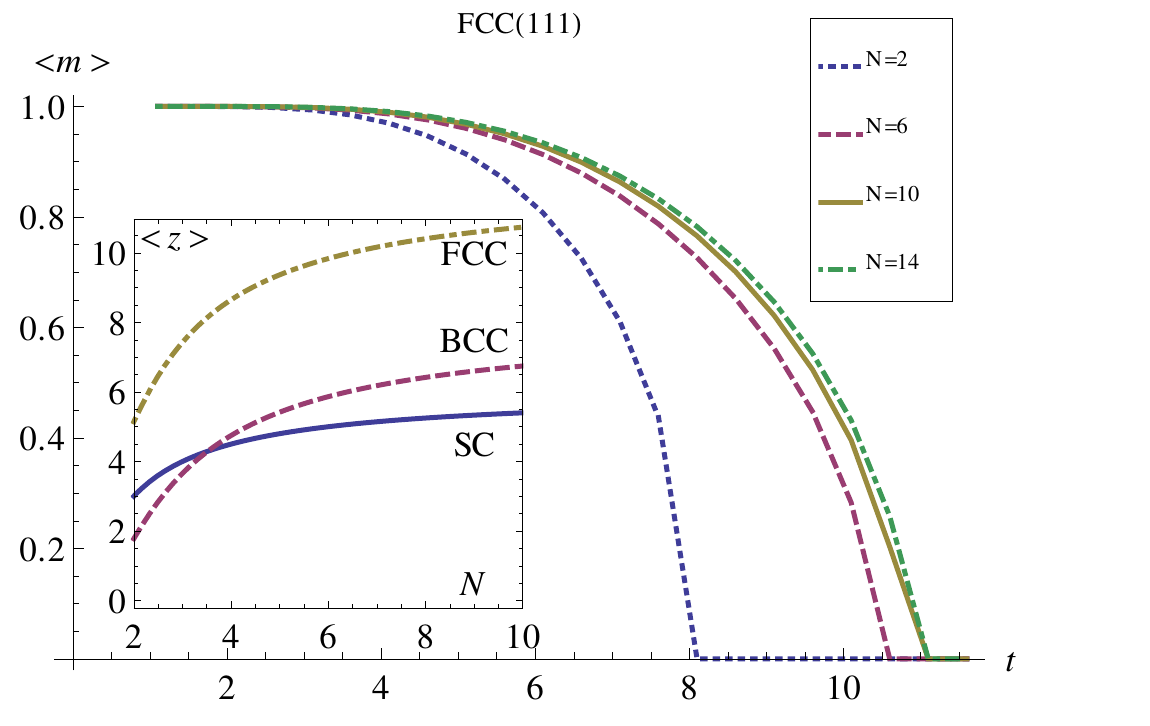}
            \includegraphics[width=8.5cm]{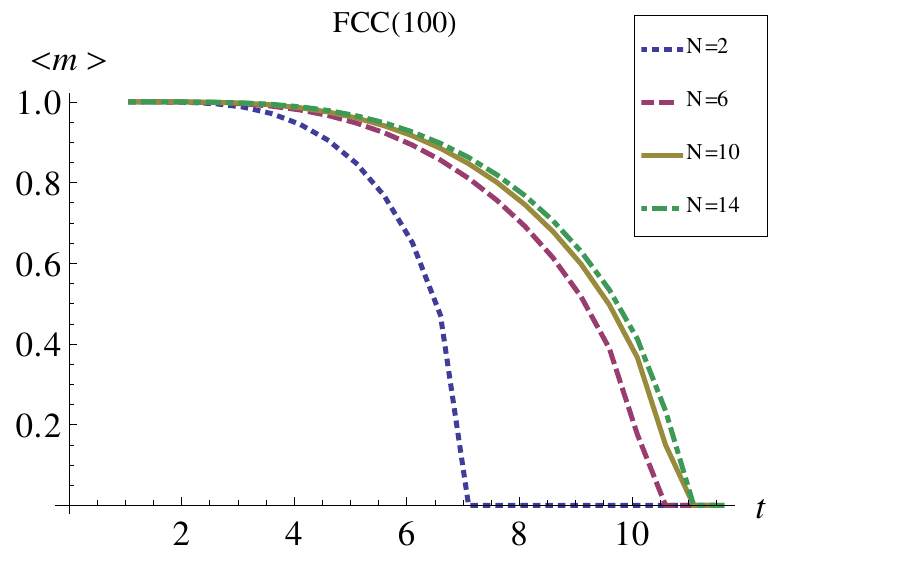}
        \end{tabular}
        \caption{The dependence of the average magnetic moment $\left\langle m\right\rangle$ on the temperature 
            for different crystallographic planes of the FCC lattice in the films of different thickness. 
            On the inset on the left it is shown an increase in the average number of 
            nearest neighbors $\left\langle z\right\rangle$
            in films of different crystalline structures, depending on the thickness of the $N$.}
        \label{fig1}
    \end{center}
    \end{figure}

Fig.~\ref{fig1} shows the temperature dependence of the average relative magnetic moment 
$\left\langle m\right\rangle =\sum^N_j{\left\langle \mu_j\right\rangle}/N$ films of different thickness. 
From the graphs it follows that the decrease in the number of monolayers leads to a reduction 
in the average number of nearest neighbors and, consequently, to lower the transition temperature~$T_c$. 
Moreover, the temperature dependence of $\left\langle m\right\rangle =\left\langle m(T)\right\rangle$, 
as well as the position of the Curie point, determined not only by the type of crystal lattice, 
but also by the crystallographic orientation of the plane of the film growth. 
The above-mentioned feature is related to the difference in the number of nearest neighbors between 
atoms in the same monolayer, and atoms located in adjacent layers.
For example, in a material with a FCC lattice grown on $(100)$-plane, each atom in the monolayer has~4 neighbors, 
and~4 more neighbors in the monolayer, which is located nearby.
When the film of the same material grows on $(111)$-plane, the number of nearest neighbors changes to~6 and~3, respectively.

\begin{figure}
    \begin{center}
    \includegraphics[width=9cm]{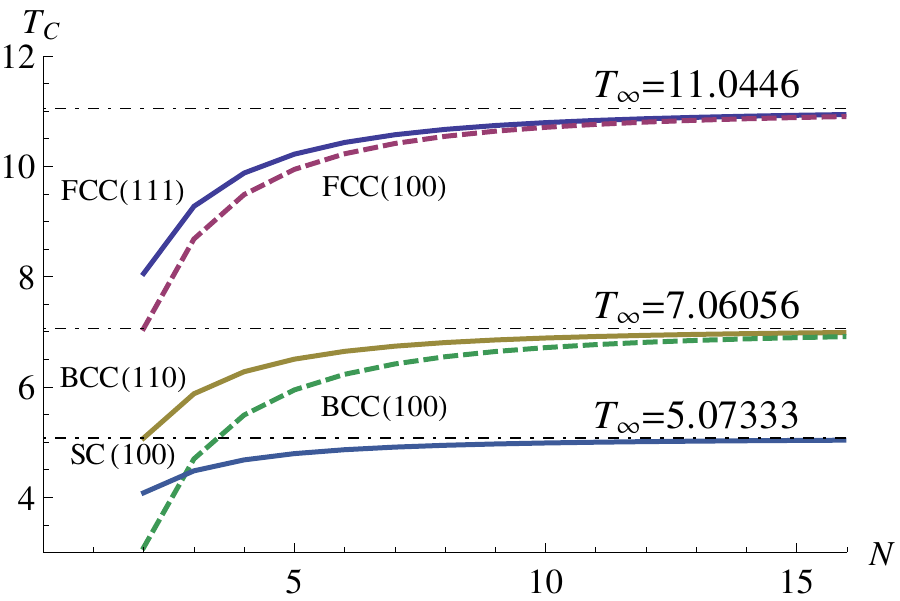}
    \caption{The dependence of the reduced Curie temperature $T_C$ on the thickness of ultrathin film 
    ($N$ is the number of monolayers) for different crystal lattices (FCC and BCC) and different 
    crystallographic orientations of the surface. Compare with the inset in Fig.~\ref{fig1}}
    \label{fig2}
    \end{center}
    \end{figure}

The dependence of the Curie temperature $T_c$ on the thickness of films with different crystal structures,
that have been calculated by solving \eqref{GrindEQ__4_}, is shown on the Fig.~\ref{fig2}.
From the illustration it is clear that $T_c$ essentially depends on the type of crystal lattice and 
is almost independent of the crystallographic orientation of the surface of the film, 
except for the area of small thickness (2 --- 6 monolayers). This effect can be explained by the fact
that in the area of small amount of monolayers the mean number of neighbors differs notably.

\begin{figure}
    \begin{center}
        \begin{tabular}{cc}
            \includegraphics[width=8.0cm]{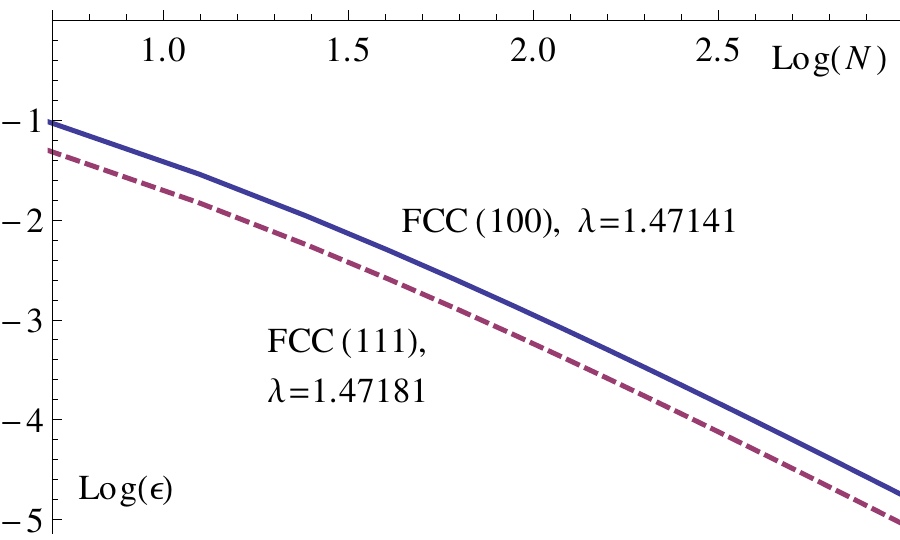}&
            \includegraphics[width=8.0cm]{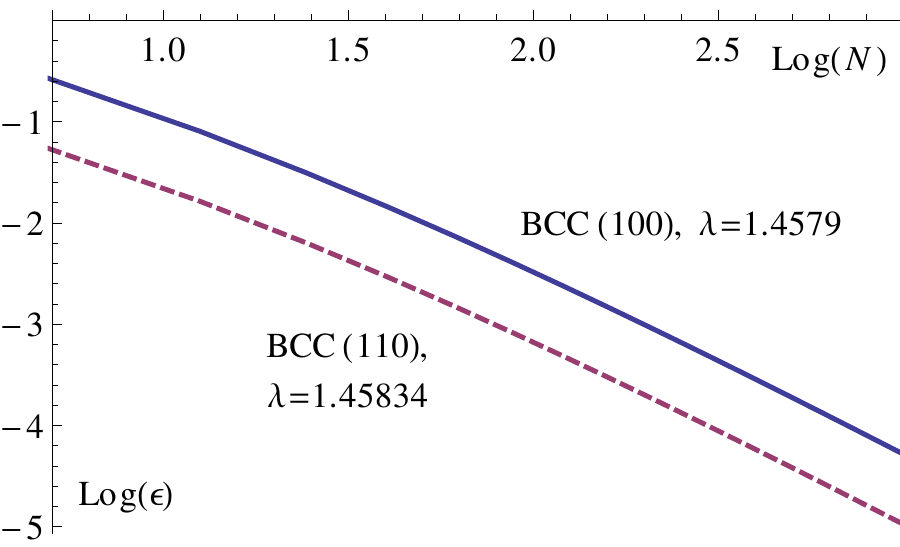}
        \end{tabular}
    \caption{The logarithmic dependence of relative change of the Curie temperature $\varepsilon (N)$
    on the film thickness $N$ (in monolayers) for FCC (on the left) and BCC lattice.}
    \label{fig3}
    \end{center}
\end{figure}

The calculation of relative change of the phase transition temperature with the growth of film thickness 
(in logarithmic scale) is shown in Fig.~\ref{fig3}.
The dependence $\varepsilon =\varepsilon (N)$ can be approximated by expression~\eqref{GrindEQ__1_}. 
The results can be compared with the experimental data (see the table).
Within the measurement errors and the accuracy of approximation, the calculated values of $\lambda$ 
and the critical exponent of the spin-spin correlation $\nu$ are close to the experimental 
that obtained on films of $Ni/Cu(111)$ and $Ni/Cu(100)$.
We note that the argument $\lambda$ does not depend on the crystallographic orientation 
of the surface ($\lambda_{110}\approx \lambda_{111}$), 
nor the type of lattice ($\lambda$ for FCC and BCC lattices differs by $0.9\,\%$).
In addition, the critical exponent of the spin-spin correlation $\left\langle \nu \right\rangle =\frac{1}{\lambda}=0.68$,
calculated in this framework, is close to the value $\nu {\rm =0.63}$, obtained by using renormalization group 
calculations in three-dimensional Ising model~\cite{Guillou1980,Guillou1977}.

\begin{table}
    \begin{center}
    {\textbf{Table} of theoretical and experimental values
    of argument $\lambda$ and the critical exponent $\nu$:\\ \vspace{5pt}}
    
    \begin{tabular}{|p{1.7in}|p{1.1in}|p{1.0in}|p{1.0in}|p{0.7in}|} \hline 
    Film & $l$ (monolayers) & $\lambda $ & $\nu $ & Link \\ \hline 
    Ni/Cu(111)  & 1 -- 8 & $1.48\pm0.20$ & 0.68$\pm$0.09 & \cite{Ballentine1989} \\ \hline 
    Ni/Cu(111) & 1 -- 10 & $1.44\pm0.20$ & 0.70$\pm$0.10 & \cite{Ballentine1990} \\ \hline 
    Calculation: FCC (111) & 2 -- 10  & 1.43 & 0.70 &  \\ \hline 
    Ni/Cu(100) & 4 -- 26 & $1.42$ & 0.70  & \cite{Schulz1994} \\ \hline 
    Calculation: FCC (100) & 2-- 26 & 1.48 & 0.68 &  \\ \hline 
    \end{tabular}
    \end{center}
\end{table}

\section{Conclusion}
The model of randomly interacting atomic magnetic moments at its relative simplicity 
provides two valuable results. Firstly, it allows to assess the influence of the thickness of ultrathin film $N$
on the temperature of magnetic phase transition, and, secondly, to establish a power-dependence of 
the relative change of the Curie temperature of $N$:
$\varepsilon (N)\sim N^{{1}/{\nu}}$, where $\nu$ is the critical exponent of the spin-spin correlation.
The obtained regularities and numerical values of critical exponents agree well with experimental 
data~\cite{Huang1993,Huang1994,Ballentine1989,Ballentine1990,Schulz1994} as well as
with RG calculations in three-dimensional Ising model~\cite{Guillou1980,Guillou1977}.
The magnetic properties of the lattice (in particular, the independence of the choice of 
crystallographic orientation of the surfaces and the type of lattice) 
are consistent with the general concepts of critical scaling.

\small
\bibliographystyle{ieeetr}
\bibliography{hongkong}

\begin{thebibliography}{10}

\bibitem{Liu1990}
C.~Liu and S.~Bader, ``{Two-dimensional magnetic phase transition of ultrathin
  iron films on Pd (100)},'' {\em Journal of Applied Physics}, vol.~67, no.~9,
  pp.~5758--5760, 1990.

\bibitem{Liu1991}
C.~Liu and S.~Bader, ``{Magnetic properties of ultrathin epitaxial films of
  iron},'' {\em Journal of Magnetism and Magnetic Materials}, vol.~93,
  pp.~307--314, 1991.

\bibitem{Rau1993}
C.~Rau, P.~Mahavadi, and M.~Lu, ``{Magnetic order and critical behavior at
  surfaces of ultrathin Fe (100) p (1$\times$ 1) films on Pd (100)
  substrates},'' {\em Journal of applied physics}, vol.~73, no.~10,
  pp.~6757--6759, 1993.

\bibitem{Qui1991}
Z.~Qiu, J.~Pearson, and S.~Bader, ``Magnetic phase transition of ultrathin fe
  films on ag(111),'' {\em Phys. Rev. Lett.}, vol.~67, no.~12, pp.~1646--1649,
  1991.

\bibitem{Li1992}
Y.~Li and K.~Baberschke, ``Dimensional crossover in ultrathin ni(111) films on
  w(110),'' {\em Phys. Rev. Lett.}, vol.~68, no.~8, pp.~1208--1211, 1992.

\bibitem{Huang1993}
F.~Huang, G.~Mankey, M.~Kief, and R.~Willis, ``{Finite-size scaling behavior of
  ferromagnetic thin films},'' {\em Journal of applied physics}, vol.~73,
  no.~10, pp.~6760--6762, 1993.

\bibitem{Huang1994}
F.~Huang, M.~Kief, G.~Mankey, and R.~Willis, ``Magnetism in the few-monolayers
  limit: A surface magneto-optic kerr-effect study of the magnetic behavior of
  ultrathin films of co, ni, and co-ni alloys on cu(100) and cu(111),'' {\em
  Phys. Rev. B}, vol.~49, no.~6, pp.~3962--3971, 1994.

\bibitem{Kohlhepp1992}
J.~Kohlhepp, H.~Elmers, S.~Cordes, and U.~Gradmann, ``Power laws of
  magnetization in ferromagnetic monolayers and the two-dimensional ising
  model,'' {\em Phys. Rev. B}, vol.~45, no.~21, pp.~12287--12291, 1992.

\bibitem{Barber1983}
M.~Barber, ``{Finite-size scaling},'' {\em {in Phase Transitions and Critical
  Phenomena}}, vol.~8, pp.~146--259, 1983.

\bibitem{Ballentine1989}
C.~Ballentine, R.~Fink, J.~Araya-Pochet, and J.~Erskine, ``Exploring magnetic
  properties of ultrathin epitaxial magnetic structures using magneto-optical
  techniques,'' {\em Applied Physics A: Materials Science $\&$ Processing},
  vol.~49, pp.~459--466, 1989.

\bibitem{Ballentine1990}
C.~Ballentine, R.~Fink, J.~Araya-Pochet, and J.~Erskine, ``Magnetic phase
  transition in a two-dimensional system: p(1$\times$1)-ni on cu(111),'' {\em
  Phys. Rev. B}, vol.~41, no.~4, pp.~2631--2634, 1990.

\bibitem{Schulz1994}
B.~Schulz, R.~Schwarzwald, and K.~Baberschke, ``{Magnetic properties of
  ultrathin Ni/Cu (100) films determined by a UHV-FMR study},'' {\em Surface
  science}, vol.~307, pp.~1102--1108, 1994.

\bibitem{Belokon2001}
V.~Belokon and K.~Nefedev, ``Distribution function for random interaction
  fields in disordered magnets: Spin and macrospin glass,'' {\em Journal of
  Experimental and Theoretical Physics}, vol.~93, pp.~136--142, 2001.
\newblock 10.1134/1.1391530.

\bibitem{Guillou1980}
J.~{Le Guillou} and J.~{Zinn-Justin}, ``{Critical exponents from field
  theory},'' {\em Phys. Rev. B}, vol.~21, pp.~3976--3998, 1980.

\bibitem{Guillou1977}
J.~Le~Guillou and J.~Zinn-Justin, ``Critical exponents for the $n$-vector model
  in three dimensions from field theory,'' {\em Phys. Rev. Lett.}, vol.~39,
  no.~2, pp.~95--98, 1977.

\end{thebibliography}
\end{document}